\documentclass{IEEEmce}
\usepackage[colorlinks,urlcolor=blue,linkcolor=blue,citecolor=blue]{hyperref}

\usepackage{hyperref}
\hypersetup{ 
 colorlinks=true, 
 linkcolor=red, 
 filecolor=blue, 
 citecolor = blue, 
 urlcolor=blue, 
 } 
 
\usepackage{upmath}

\jvol{XX}
\jnum{XX}
\paper{XX}
\jmonth{xxx/xxx}
\publisheddate{DD MM YYYY}
\currentdate{DD MM YYYY}
\jname{IEEE Security \& Privacy Magazine}
\pubyear{YYYY}
\doiinfo{MCE.YYYY.Doi Number}

\begin{document}

%%%%
% NOTE: Add the Running Head Title below!
%%%%
\sptitle{TitanCA project} 

% \title{The Road to 100+ CVEs: Orchestrating LLM-Powered Agents for Vulnerability Discovery}
% \title{TitanCA: An Orchestrated Multi-Agent Framework for Automated Vulnerability Discovery in Large-Scale Codebases}
\title{TitanCA: Lessons from Orchestrating LLM Agents to Discover 100+ CVEs}

%%%%
% NOTE: Authors from the same institution in the following sequence must be listed on a single line! Do NOT indicate the Corresponding Author behind the authors' name. Any acknowledgment must go after the conclusion section.
%%%%

\author{Ting Zhang}
\affil{Monash University}

\author{Yikun Li, 
Chengran Yang, 
Ratnadira Widyasari, 
Yue Liu, 
Ngoc Tan Bui, 
Phuc Thanh Nguyen, 
Yan Naing Tun,
Ivana Clairine Irsan, 
Huu Hung Nguyen, 
Huihui Huang, 
Jinfeng Jiang,
Lwin Khin Shar, 
Eng Lieh Ouh, 
David Lo}
\affil{Singapore Management University}

\author{Hong Jin Kang}
\affil{The University of Sydney}

\author{Yide Yin, Wen Bin Leow}
\affil{GovTech, Singapore}

%%%%
% NOTE: Add the Running Head and Article Titles below.
%%%%

\markboth{TitanCA}{An Orchestrated Multi-Agent Framework}

\begin{abstract}
Software vulnerabilities remain one of the most persistent threats to modern digital infrastructure. 
While static application security testing (SAST) tools have long served as the first line of defense, they suffer from high false-positive rates. 
This article presents TitanCA, a collaborative project between Singapore Management University and GovTech Singapore that orchestrates multiple large language model (LLM)-powered agents into a unified vulnerability discovery pipeline. 
Applied in open-source software, TitanCA has discovered 203 confirmed zero-day vulnerabilities and yielded 118 CVEs.
We describe the four-module architecture, i.e., matching, filtering, inspection, and adaptation, and share key lessons from building and deploying an LLM-based vulnerability discovery solution in practice.

\end{abstract}

\maketitle

\enlargethispage{10pt}

\chapterinitial{T}his article offers a synthesis of the studies underpinning TitanCA~\cite{bui2025vulcoco,weyssow2025r2vul,vultrial,li2025out,li2024cleanvul} along with our opinions, presented at an accessible level with full technical detail deferred to the corresponding papers where available.
The discovery of software vulnerabilities before they can be exploited is a central challenge in cybersecurity. 
Despite decades of research in static analysis, fuzzing, and formal verification, the volume and complexity of modern codebases continue to outpace traditional detection methods. 
Popular static application security testing (SAST) tools, while widely adopted in continuous integration and continuous deployment (CI/CD) pipelines, produce substantial numbers of false positives that hurt developer trust and consume triage resources. 
The consequence is that developers begin ignoring security warnings entirely, leaving many vulnerabilities untriaged.

Recent advances in large language models (LLMs) have opened new avenues for automated code understanding. 
Models trained on billions of lines of source code can reason about code semantics and identify vulnerable patterns that rule-based analyzers miss. 
Yet applying LLMs to vulnerability detection in a new production environment introduces its own challenges: individual models are prone to hallucination, precision matters more than recall when developer attention is scarce, and general-purpose models lack awareness of the vulnerability patterns specific to a given deployment context.

TitanCA addresses these challenges through orchestration rather than by scaling models. 
Rather than relying on a single model, the system composes multiple LLM-powered agents into a four-module pipeline in which each stage refines the output of the previous one, progressively filtering false positives and increasing confidence in the findings. 
The final module incorporates domain knowledge drawn from the target organization to capture patterns that generic models overlook.

TitanCA is a collaboration between Singapore Management University (SMU) and the Government Technology Agency of Singapore (GovTech). 
The work described in this article corresponds to Phase 1 of the TitanCA project, which concluded in January 2026.
The full pipeline has been delivered to the target organization, and a lightweight version has been deployed. 
We also run the first three modules (without the domain-specific adaptation module) in a continuous fashion to monitor a wide variety of OSS repositories.
As of March 2026, TitanCA has analysed code from over 127,000 GitHub repositories, identified 203 zero-day vulnerabilities, all subsequently remediated, and published 118 CVEs as a direct result of our approach.
This article describes the technical architecture of TitanCA, presents deployment results on OSS, and distils engineering lessons that may guide other teams building LLM-powered security tooling.

\section{How TitanCA works}
\begin{figure*}
    \centering
    \includegraphics[width=\linewidth]{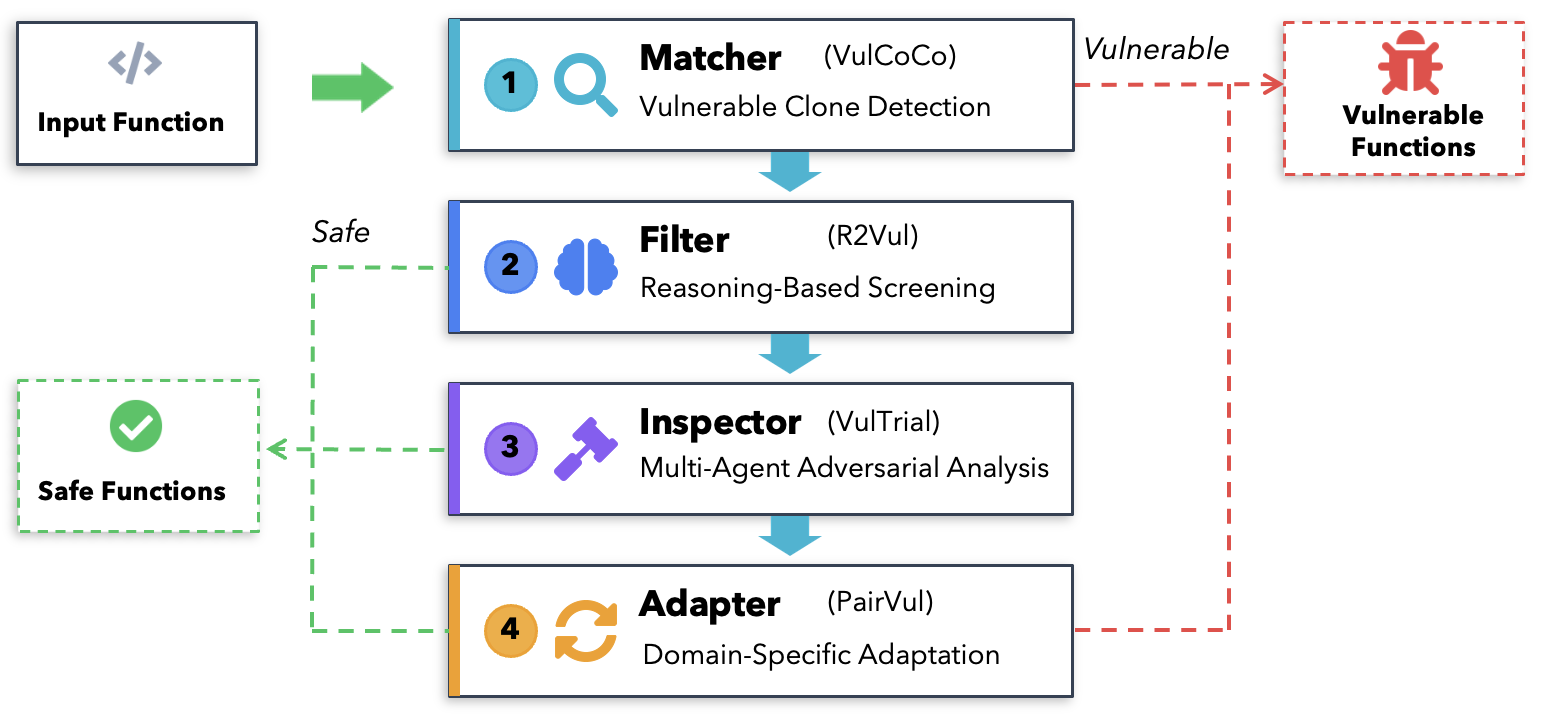}
    \caption{Overview of the TitanCA pipeline. The four modules progressively refine detection results: the Matcher (VulCoCo) casts a wide net for candidate vulnerabilities, the Filter (R2Vul) reduces false positives through structured reasoning, the Inspector (VulTrial) applies multi-agent deliberation, and the Adapter (PairVul) handles domain-specific adaptation.}
    \label{fig:overview}
\end{figure*}

TitanCA implements a just-in-time (JIT) vulnerability-discovery approach that analyzes code as it is committed, rather than during periodic audits. 
When a developer pushes a commit to the platform, the system extracts the changed code and passes it through a four-module pipeline. 
These modules are designed to complement one another, creating a layered defense that prioritizes precision while also maintaining recall.
Figure~\ref{fig:overview} shows the overview of the TitanCA pipeline.

\subsection{Module 1: The Matcher (VulCoCo)}
The first module performs vulnerable clone detection. 
VulCoCo~\cite{bui2025vulcoco} generates embeddings of the candidate functions and compares them against a curated database of known vulnerable functions.
Our main sources are the NVD and the TitanVul dataset~\cite{li2025out}, which contains approximately 40,000 vulnerability samples spanning diverse languages and vulnerability types. 
The module uses similarity search over a vector database to identify candidate matches, then employs an LLM as a semantic validator to confirm whether the detected similarity reflects a genuine vulnerability pattern rather than a superficial syntactic resemblance. 
If the candidate function is flagged as vulnerable, we skip the remaining modules; otherwise, the function flows to the following modules.
% This initial scan maximizes recall at the expense of precision.
This initial scan maximizes recall because the similarity matches are highly likely to be genuine vulnerabilities.
This is a deliberate design choice, since subsequent modules are more computationally expensive while contributing more towards false negative reduction.
For the complete clone-detection algorithm and its evaluation, see Bui et al.~\cite{bui2025vulcoco}. 

\subsection{Module 2: The Filter (R2Vul)}
The second module applies reasoning-based screening to further identify true vulnerabilities among the candidates that the Matcher did not flag, i.e., functions that did not match any known vulnerability pattern but may still contain novel or previously unseen weaknesses.
R2Vul~\cite{weyssow2025r2vul} is trained with structured reasoning distillation that distinguishes between two types of model reasoning: grounded reasoning, where the logical chain matches the label, and misleading reasoning, where the explanation sounds plausible but is logically disconnected from the label. 
The training process uses contrastive supervision with RLAIF (Reinforcement Learning from AI Feedback), presenting the model with paired examples of grounded and misleading analyses of the same code fragment via the ORPO preference optimization objective. 
This teaches the model not merely to classify code as vulnerable or safe, but to produce an explicit chain of reasoning that justifies its judgment.

R2Vul also incorporates a lightweight calibration step that leverages the model's own reasoning structure. 
At inference time, the system compares the conditional log-likelihoods of the model generating reasoning under the vulnerable versus safe template prefixes, converting the log-odds margin into a confidence score through a logistic mapping. 
A function is flagged as vulnerable only when this confidence exceeds a tunable threshold. 
In practice, this calibration reduces the false positive rate from 28\% to 20\% under balanced conditions while preserving over 77\% recall, and achieves the same false positive target even under extreme class imbalance (1:10 vulnerable-to-safe ratio) typical of production environments. 
If the candidate function is classified as safe, we exit the prediction pipeline; otherwise, it flows to Module 3 for further investigation.
Additional details on the structured reasoning distillation, RLAIF training, and calibration procedure can be found in Weyssow et al.~\cite{weyssow2025r2vul}.

\subsection{Module 3: The Inspector (VulTrial)}
The third module introduces multi-agent deliberation using a mock-courtroom metaphor. 
% VulTrial~\cite{vultrial} convenes five specialized agents, each playing a distinct role in the adjudication process. 
VulTrial~\cite{vultrial} convenes four specialized agents, each playing a distinct role in the adjudication process. 
A Security Researcher presents the case that the code is vulnerable, marshaling evidence from the earlier pipeline stages. 
A Code Author defends the code, arguing that the flagged pattern is intentional or safe. 
% A Moderator manages the proceedings, posing clarifying questions and ensuring that both sides address the strongest arguments of their opponent.
A Moderator serves as the judge, %, rather than taking sides, 
it distills the exchange into an impartial, fact-focused summary of the key points raised by both the security researcher and the code author.
% A Review Board of three domain experts evaluates the arguments presented by both sides. Finally, a Synthesizer integrates all perspectives into a final verdict with a confidence assessment.
A Review Board functions as the jury and issues the final verdict.
This adversarial structure forces the system to consider both sides of each case, mimicking the deliberative process that experienced security analysts employ during manual code review. 
The multi-agent design helps surface subtle vulnerabilities that a single model might miss, while filtering out false alarms.
In the OSS pipeline, this module is the final decision point. 
In the target deployment, functions classified as safe exit the pipeline here; those still flagged as vulnerable flow to Module~4 for further investigation.
The full multi-agent design and ablation results are reported in Widyasari et al.~\cite{vultrial}.

\subsection{Module 4: The Adapter (PairVul)}
The fourth module addresses the problem of domain-specific adaptation. 
When TitanCA is deployed in a new organizational context, with different programming languages, frameworks, coding conventions, and vulnerability patterns, the coding conventions and existing vulnerabilities may differ substantially from the training data. 
To address this gap, we first deployed the lightweight version in the target organization to get initial results.
To ensure the high quality of the labels, we first manually labeled a sample of functions, then applied LLM-based relabeling.
We propose PairVul, which analyzes the false positives generated during initial deployment to identify recurring error patterns.
PairVul then fine-tunes the detection models using these newly labeled examples. 
This creates a feedback loop that enables the system to continuously improve its precision within a specific deployment environment without requiring expensive manual annotation of new training data.

\subsection{Data Foundation and Infrastructure}
Effective LLM-based vulnerability detection requires high-quality training data at scale, and constructing such datasets is a research challenge on its own.
The TitanCA project invested heavily in data engineering, building the dataset from both real-world vulnerability disclosures and synthetic generation techniques. 
The dataset now contains over 342,000 likely vulnerability samples.
We propose CleanVul~\cite{li2024cleanvul} to address the label noise, which is a well-known but often overlooked problem in vulnerability datasets.
Many publicly available datasets contain mislabeled samples, code labeled as vulnerable when it is not, or vice versa, that silently degrade model performance during training. 
CleanVul applies systematic cleaning procedures, combining automated heuristics with LLM-assisted verification, to produce more reliable training sets. 
More recently, we propose TitanVul and BenchVul~\cite{li2025out}: a large-scale, high-quality training dataset and a manually curated benchmark covering the MITRE Top 25 Most Dangerous CWEs, revealing that in-distribution performance is a poor predictor of real-world generalization.
The operational infrastructure is substantial: approximately 500 terabytes of data engineering workloads support the monitoring of over 127,000 GitHub repositories.
This scale is essential for training models that generalize across the diversity of real-world code patterns, languages, and vulnerability types.
For the dataset construction, cleaning heuristics, and benchmark composition, see Li et al.~\cite{li2024cleanvul} and Li et al.~\cite{li2025out}.

\section{Supporting the open source community}
As of March 2026, the system has identified 203 vulnerabilities in OSS that were subsequently confirmed and fixed by their respective development teams. 
Of these, 118 have been assigned CVE identifiers and published.
The complete CVE list is publicly available in \url{https://titancaproject.github.io/cves.html}.
The severity profile of discovered vulnerabilities underscores the system's ability to detect major issues: 35\% are rated critical severity, 95\% are at least medium severity, and 91\% exhibit low attack complexity, meaning they are relatively straightforward to exploit once discovered.

Approximately half of the discovered vulnerabilities fall within the CWE Top-25 Most Dangerous Software Weaknesses, a list maintained by MITRE that represents the most prevalent and impactful vulnerability categories. 
The most frequently identified types include CWE-787 (out-of-bounds write), CWE-125 (out-of-bounds read), CWE-190 (integer overflow or wraparound), CWE-119 (improper restriction of operations within the bounds of a memory buffer), CWE-835 (loop with unreachable exit condition), and CWE-22 (improper limitation of a pathname to a restricted directory).

\section{Lessons learned}
\subsection{Orchestration over monolithic models.} 
The most important architectural insight from TitanCA is that a pipeline of specialized, collaborating agents outperforms a single model.
Each module addresses a specific failure mode: the Matcher ensures broad coverage, the Filter enforces LLMs' reasoning, the Inspector solicits opinions from multiple LLMs, and the Adapter handles distribution shift.
This division of labor is similar to how human security teams operate:  different specialists contributing complementary perspectives to a shared objective.

We did not arrive at this design on the first attempt. 
Our earliest effort tried to fine-tune a single LLM to perform end-to-end vulnerability detection. 
It achieved reasonable recall but generated an unmanageable volume of false positives.
As the project progressed, we decomposed the task into specialized stages, realizing that vulnerability detection is not one problem but several.
Those can include pattern recognition, logical reasoning, adversarial validation, and domain adaptation.
Each demands a different model capability. 
We anticipate that this observation can generalize, especially for complex security tasks, as orchestrating specialists will consistently outperform scaling a single generalist.

\subsection{Cost-aware pipeline ordering.}
The ordering of the pipeline matters as well. 
Our first design placed the multi-agent Inspector and the lightweight Filter in parallel, because the two components originated from different research motivations and were initially developed as alternative approaches. 
This turned out to be wasteful: the Inspector involves multiple LLM calls per function and is computationally expensive. 
Restructuring the pipeline so that cheap filtering runs first and expensive deliberation runs later reduced our per-function cost substantially. 
The lesson is that pipeline design for LLM-based systems should follow a similar principle as query optimization: push the cheapest, most selective operations to the front.

\subsection{Precision is critical in practice.}
In production environments, false positives are more operationally damaging than false negatives. 
Developers can be overwhelmed by an excess of warnings quickly and then tend to ignore the tool entirely.
In this project, we focus on the F0.3 metric, which weights precision three times more than recall, and better captures this operational reality than the standard F1 score. 
TitanCA's architecture is explicitly designed to maximize this metric through successive filtering stages.

We believe the academic community's reliance on F1 as the primary evaluation metric for vulnerability detection is misleading. 
F1 assumes that false positives and false negatives are equally costly, an assumption that does not hold in practical usage.
% production environments.
We encourage future work to report F0.3 or similar precision-weighted metrics alongside F1 to provide a more realistic picture of practical utility.

\subsection{Structured reasoning improves reliability.}
The R2Vul module trains LLMs to produce grounded reasoning chains rather than simply producing classification labels. 
This approach significantly improves the reliability and interpretability of model outputs. 
By contrasting grounded and misleading reasoning during training, the model learns to avoid superficially plausible but logically disconnected explanations.

An under-appreciated aspect of this problem is that a correct prediction accompanied by a wrong explanation can be more damaging than a direct incorrect prediction. 
When the model flags a function as vulnerable but cites an irrelevant reason, the developer reads the explanation, recognizes it as flawed, and drops the entire warning. 
We found that grounding the reasoning chain not only improved classification accuracy but, more importantly, reduced the rate at which developers drop valid alerts due to unconvincing explanations.

\subsection{Multi-agent debate catches what single models miss.}
The mock courtroom design of VulTrial addresses a limitation of single-model inference: confirmation bias. 
When a single model produces a vulnerability prediction, it has no internal mechanism to challenge its own reasoning. 
The multi-agent adversarial structure forces explicit consideration of counter-arguments and alternative explanations. 
This process produces more robust judgments than any single model alone.

\subsection{Adaptation is essential for deployment.}
Since no model trained exclusively on public data can perfectly capture the vulnerability patterns, coding idioms, and framework usage of a specific organization, it is critical to incorporate domain characteristics.
We initially applied only the first three modules, without the adapter, in the target environment. 
But after a few sample runs, we observed some unique false positive patterns and realized the need to learn domain-specific code patterns.
To handle these cases, we propose PairVul to systematically analyze deployment-time false positives to generate new training signals with the feedback loop.
This step is critical for maintaining and improving precision as the system encounters novel codebases and evolving software practices.
We believe these patterns need to be adapted when applied in different target organizations.

\section{The road ahead}
Phase~1 of TitanCA operated at the function level and focused exclusively on detection. 
Phase~2, which commenced in March 2026, extends the system in three directions. 
First, it expands the analysis context beyond individual functions to leverage the broader security reasoning capabilities of agentic frameworks. 
Second, it provides repair suggestions to help developers fix discovered vulnerabilities more efficiently~\cite{yang2025semantics}. 
Third, it incorporates research into developer behavior, studying how developers perceive and respond to AI-generated vulnerability reports, with the goal of improving the system's reporting interface and explanation quality.

Both phases focus on detecting vulnerabilities in developer-written code. Two complementary questions remain open: whether LLM agents can generate secure code in the first place~\cite{chen2025secureagentbench}, and whether they can effectively detect vulnerabilities in AI-generated code as LLM-based coding agents become widely adopted.

\section{Conclusion}
In TitanCA Phase 1, our experience demonstrates that orchestrating multiple LLM-powered agents into a structured, layered pipeline can achieve vulnerability discovery results that substantially exceed those of conventional SAST tools. 
As LLMs continue to improve in both capability and efficiency, systems like TitanCA point toward a future where automated vulnerability discovery operates as an effective and trustworthy complement to human security expertise.

\section*{Acknowledgment}
This research / project is supported by the National Research Foundation, Singapore and Ministry of Digital Development \& Information under its Smart Nation and Digital Government Translational R\&D Grant (Award No: TRANS2026-TGC01). Any opinions, findings and conclusions or recommendations expressed in this material are those of the author(s) and do not reflect the views of National Research Foundation, Singapore.

%%%%
% NOTE: Do not cite arXiv! Instead, give the journal titles, including vol., no., pp., and doi.
%%%%
\bibliographystyle{IEEEtran}
\bibliography{main}

\begin{IEEEbiography}{Ting Zhang} {\,} is a Lecturer in the Department of Software Systems and Cybersecurity at Monash University in Clayton, Victoria 3800, Australia. Her research interests include vulnerability detection, vulnerability repair, and secure code generation. Zhang received her Ph.D. in Computer Science from Singapore Management University. Contact her at \url{https://happygirlzt.com/academic} or \url{happygirlzt@gmail.com}.
\end{IEEEbiography}

\begin{IEEEbiography}{Yikun Li} {\,} is a Research Scientist at Singapore Management University. His research interests include software vulnerability detection and repair, and AI-driven software security. Li received his PhD in Computer Science from the University of Groningen, the Netherlands. Contact him at \url{yikunli@smu.edu.sg}.
\end{IEEEbiography}

\begin{IEEEbiography}{Chengran Yang} {\,} is a Research Scientist at Singapore Management University. His research interests include trustworthy code models, code vulnerability analysis, and code testing. Yang received his PhD in computer science from Singapore Management University. Contact him at \url{chengran98@gmail.com}.
\end{IEEEbiography}

\begin{IEEEbiography}{Ratnadira Widyasari} {\,} is a Research Scientist at Singapore Management University. Her research focuses on advancing software quality assurance through automation and explainability. Widyasari received her Ph.D. from Singapore Management University. Contact her at \url{https://ratnadiraw.github.io/}.
\end{IEEEbiography}

\begin{IEEEbiography}{Yue Liu} {\,} is a Research Scientist at Singapore Management University. His research interests include trustworthy AI for software development, secure AI coding tools, and developer trust in AI-assisted programming. Liu received his Ph.D. in information technology from Monash University. Contact him at \url{https://yueyuel.github.io/}.
\end{IEEEbiography}

\begin{IEEEbiography}{Ngoc Tan Bui} {\,} is a PhD student at Singapore Management University. His research interests include vulnerability detection, vulnerability repair, and patch validation. Bui received his bachelor's degree in computer science from Hanoi University of Science and Technology. Contact him at \url{ngoctanbui@smu.edu.sg}.
\end{IEEEbiography}

\begin{IEEEbiography}{Phuc Thanh Nguyen} {\,} is a Research Engineer at Singapore Management University. His research interests include vulnerability detection and software security topics. Nguyen received his bachelor's degree in data science from Hanoi University of Science and Technology, Vietnam. Contact him at \url{ptnguyen@smu.edu.sg}.
\end{IEEEbiography}

\begin{IEEEbiography}{Yan Naing Tun} {\,} is a Research Engineer at Singapore Management University. His research interests include software testing and security, vulnerability detection, Android and IoT security, and AI-assisted software engineering. Tun received his B.Sc. in Computer Science from the University of Computer Studies, Mandalay, Myanmar. Contact him at \url{yannaingtun@smu.edu.sg}.
\end{IEEEbiography}

\begin{IEEEbiography}{Ivana Clairine Irsan} {\,} is a PhD student at Singapore Management University. Her research interests include vulnerability detection and software security topics. 
Irsan received her master's degree in Computer Science from Institut Teknologi Bandung. Contact her at \url{ivanairsan@smu.edu.sg}.
\end{IEEEbiography}

\begin{IEEEbiography}{Huu Hung Nguyen} {\,} is a PhD student at Singapore Management University. His research interests include vulnerability detection and software supply chains. Nguyen received his bachelor's degree in Information Technology from Hanoi University of Science and Technology. Contact him at \url{huuhungn@smu.edu.sg}.
\end{IEEEbiography}

\begin{IEEEbiography}{Huihui Huang} {\,} is a PhD student at Singapore Management University. Her research interests include AI for software engineering, vulnerability detection, and agentic penetration testing. Huang received her bachelor's degree in computer science and technology from Southern University of Science and Technology. She is a student member of ACM. Contact her at \url{hhhuang@smu.edu.sg}.
\end{IEEEbiography}

\begin{IEEEbiography}{Jinfeng Jiang} {\,} is a Research Engineer at Singapore Management University. His research interests include AI for Software Engineering and Trustworthy Code LLM. Jiang received his Master's degree in Software Engineering from Tongji University. Contact him at \url{jfjiang@smu.edu.sg}.
\end{IEEEbiography}

\begin{IEEEbiography}{Lwin Khin Shar} {\,} is an Associate Professor of Computer Science (Practice) at Singapore Management University. His research interests include security testing and analysis of web/mobile applications and cyber physical systems. Shar received his Ph.D in Software Engineering from Nanyang Technological University, Singapore. He is a member of IEEE and ACM. Contact him at \url{lkshar@smu.edu.sg}.
\end{IEEEbiography}

\begin{IEEEbiography}{Eng Lieh Ouh} {\,} is an Associate Professor of Computer Science (Education) at Singapore Management University. His research interests include AI-enabled vulnerability detection, secure software engineering, and computing education. Ouh received his Ph.D in Computer Science from the National University of Singapore. He is a Senior Member of IEEE and an ISC2 member and authorised instructor. Contact him at \url{elouh@smu.edu.sg}.
\end{IEEEbiography}

\begin{IEEEbiography}{David Lo} {\,} is the OUB Chair Professor of Computer Science and Vice Provost (Research) Designate at Singapore Management University. His research interests include software engineering, AI, and cybersecurity. Lo received his PhD in Computer Science from National University of Singapore. He is an ACM Fellow, IEEE Fellow, Fellow of Automated Software Engineering, and NRF Investigator (Senior Fellow). Contact him at \url{https://faculty.smu.edu.sg/profile/david-lo-901} or \url{davidlo@smu.edu.sg}.
\end{IEEEbiography}

\begin{IEEEbiography}{Hong Jin Kang} {\,} is a Lecturer in the School of Computer Science at The University of Sydney in Sydney, New South Wales 2006, Australia. His research interests include program analysis and secure software engineering. Kang received his PhD in Computer Science from Singapore Management University. Contact him at \url{kanghj.github.io} or \url{hongjin.kang@sydney.edu.au}.
\end{IEEEbiography}

\begin{IEEEbiography}{Yin Yide} {\,} is a Cybersecurity Engineer at the Government Technology Agency of Singapore. His research interests include vulnerability detection and automatic patch generation. Yide received his MSc in Computing Science from Imperial College London. Contact him at \href{mailto:yin_yide@tech.gov.sg}{yin\_yide@tech.gov.sg}.
\end{IEEEbiography}

\begin{IEEEbiography}{Leow Wen Bin} {\,} is a Cybersecurity Engineer in the Government Technology Agency of Singapore. His research interests include vulnerability detection, secure code generation, and federated learning. Leow received his Master of Science in Engineering from Tsinghua University. Contact him at \url{linkedin.com/in/leowwb} or \href{mailto:leow_wen_bin@tech.gov.sg}{leow\_wen\_bin@tech.gov.sg}.
\end{IEEEbiography}

\end{document}